\documentclass[referee]{cjaa}
\usepackage{graphicx}
\input{epsf.sty}                        %for PS/EPS graphics inclusion, old
\input{psfig.sty}
\begin{document}

\title{Spectroscopic Study of SU UMa-type Dwarf Nova YZ~Cnc during its 2002 Superoutburst}
\author{Ying-he Zhao, Zong-yun Li\mailto{}, Xiao-an Wu and Qiu-he Peng}

\institute{Department of Astronomy, Nanjing University, Nanjing,
210093, China\\
\email{zyli@nju.edu.cn}}

\date{Accepted ~, Received ~}

\abstract{ We report time-resolved spectroscopic observations of
the SU~Ursae Majoris dwarf nova, YZ Cnc, for 2 nights over 11 hrs
during its 2002 January superoutburst. The spectra only show
absorption-line profiles in the first day. But the lines display
blue and red troughs, with ``W'' profiles in the second day. The
radial velocity curve of the absorption troughs and emission peaks
of H$\beta$ has an amplitude of $49\pm10$ km s$^{-1}$ and a phase
offset of $-0.07\pm0.04$, which are very similar to those measured
in quiescence; however, the $\gamma$ velocity deviates strongly
from the systemic velocity measured in quiescence, showing
variation of the order of $\pm$60 km s$^{-1}$. And large shifts of
$\sim$70 km s$^{-1}$ and  $\sim$0.09, for the orbital-averaged
velocity and phase respectively, are also found in our
observations. All these phenomena can be well explained with a
precession of an eccentric disk and we conclude that these
phenomena are the characteristic products of an eccentric
accretion disk.  }

\titlerunning{Spectroscopic study of YZ Cnc during superoutburst}
\authorrunning{Y. H. Zhao, Z. Y. Li, X. A. Wu et al.}
\maketitle

\section{Introduction}
YZ Cnc is a member of the class of SU UMa type dwarf novae, in
which normal outbursts (i.e., short outbursts) are occasionally
interspersed by longer and brighter distinctive superoutbursts,
accompanying superhump phenomena. Superhumps are large amplitude
luminosity variations with a period usually a few percent longer
than the orbital period of the binary system. And this is
generally thought to arise from the interaction of the donor star
orbit with slowly progradely precessing non-axisymmetric accretion
disk. The eccentricity of the disk arises because a 3:1 resonance
occurs between the donor star orbit and motion of matter in the
outer disk (for a good review, see Warner 1995).

YZ Cnc is remarkable in several respects. Photometric study showed
that YZ Cnc has a visual magnitude of $\sim$14.5 in quiescence and
$\sim$10.5 in outburst and is one of the most active cataclysmic
variables because of the large flickering amplitude of 0.75 mag
peak to peak (Moffett \& Barnes 1974). It has a very short
recurrence time of $\sim$11.3 days (Vorob'yeva \& Kukarkin 1961).
Patterson (1979) discovered superhumps in the light curve with a
period of 0.09204 day of YZ Cnc and defined it as an SU UMa type
star, and his continuous study (Patterson 1981) with high-speed
photometric observations of YZ Cnc found no evidence for coherent
oscillations either in quiescence or during eruptions. But van~
Paradijs et al. (1994) found that orbital variability was present
during quiescence.

In X-ray band, YZ Cnc has also been studied very extensively with
the Einstein satellite (C\'{o}rdova \& Mason 1984, Eracleous et
al. 1991), with EXOSAT (van der Woerd 1987), with the ROSAT PSPC
during the ROSAT All Sky Survey and in subsequent pointings
(Verbunt et al. 1997, 1999; van Teeseling \& Verbunt 1994), and
with XMM-Newton (Hakala et al. 2004).

However, spectroscopy has not been as extensive as photometry or
X-ray, specially when the system is undergoing superoutburst. The
sporadical spectroscopic observations have been performed by
several authors as part of general surveys of cataclysmic
variables (Szkody 1981, Oke \& Wade 1982, Wade 1982, Williams
1983). Shafter \& Hessman (1988, named SH hereafter) presented a
detailed spectroscopic study of YZ Cnc when the star was in
quiescence and gave an orbital period of 0.0868(2) day. According
to this orbital period and the superhump period given by Patterson
(1979), a precessing period of 1.52 day can be obtained. We thus
did a 2-days observation to study the accretion disk of YZ Cnc
during its 2002 January superoutburst. In this paper we report our
observations and reduction of the spectroscopic data in Sect. 2.
In Sect. 3 we describe the main characteristics of the
spectroscopic results and explanations for these results. In the
last two sections we present a brief discussion and conclusions
for this work.

\section{Observations}
The observations were conducted with the Optomechanics Research,
Inc., Cassegrain spectrograph attached to the 2.16-m telescope
with a TEK1024 CCD camera at Xinglong Station of the National
Astronomical Observatory. Total observational time was 11 hrs, 5.3
times of the orbital period. A 300 groove mm$^{-1}$ grating blazed
at 5000 \AA\ was used, and the slit width was set to 2$''$.5. Dome
flats were taken at the beginning and end of each night. Exposure
time for the star ranged from 600 to 1800, depending on weather
conditions. Fifteen and fourteen star spectra were collected on
January 21 and 22 (Beijing time), respectively. The journal of the
observations is listed in Table 1.

The technique of data processing is similar to that in Wu et al.
(2001). After bias subtraction and flat field correction, we used
the $IRAF$\footnotemark, \footnotetext{IRAF is distributed by the
National Optical Astronomy Observatories, which is operated by
Associated of Universities for Research in Astronomy, Inc., under
cooperative agreement with the National Science Foundation.} task
$cosmicray$ to eliminate the cosmic rays roughly and then used
$imedit$ to get the cosmic rays rejected more clearly by hand. The
lamp spectra recorded before and after every two successive star
exposures were used to interpolate the coefficients of the
wavelength scales. We derived a spectral resolution of 12 \AA \
from FWHM measurement of the lamp spectra. The rms error of
identified lines was less than 0.2 \AA \ using a fourth-order
Legendre polynomial to fit the lines, corresponding to 12 km
s$^{-1}$ near H$\beta$. The flux was calibrated used the standard
star, HD109995, and had an estimate error of $\sim$10\%.

%-------------------Table 1------------------------------
\begin{table}
\caption{Journal of observations.}
\begin{tabular}{ccccc}
\hline\hline
 Date (UT) & HJD Start & Duration & Exposure & Plates\\
(Year 2002) & -2452000 & (hr)     & (s)     &   \\
\hline
Jan 21 ........& 296.1457 & 5.22 & 900,1200 & 15\\
Jan 22 ........& 297.1064 & 5.78 & 1500 & 14\\
\hline
\end{tabular}
\end{table}

\section{Results and Analysis}
\subsection{Average Spectra}
Fig. 1 shows the average spectra of YZ Cnc during its
superoutburst. The top and lower panel are the sum of all 15
individual spectra recorded on January 21 and the sum of all 14
individual spectra obtained on January 22, respectively. The
spectra of January 21 is characterized by broad Balmer absorption
and an energy distribution significantly bluer than that of
January 22, when the eruption was fading out. This spectra is
typical of an optically thick accretion disk with a high accretion
rate. The emission component came out and was specially stronger
in H$\beta$ absorption on January 22. The He~I ($\lambda
\lambda$4471, 4922) absorption, He~I $\lambda$5015+Fe~II
$\lambda$5018 and Fe~II $\lambda$5169 absorption (emission on
January 22) are also present (see more clearly in Fig.~2). The
equivalent widths of the absorption and emission lines are
summarized in Table 2.

%----------------------Figure 1-------------------------
\begin{figure}
\centering
\includegraphics[width=13cm]{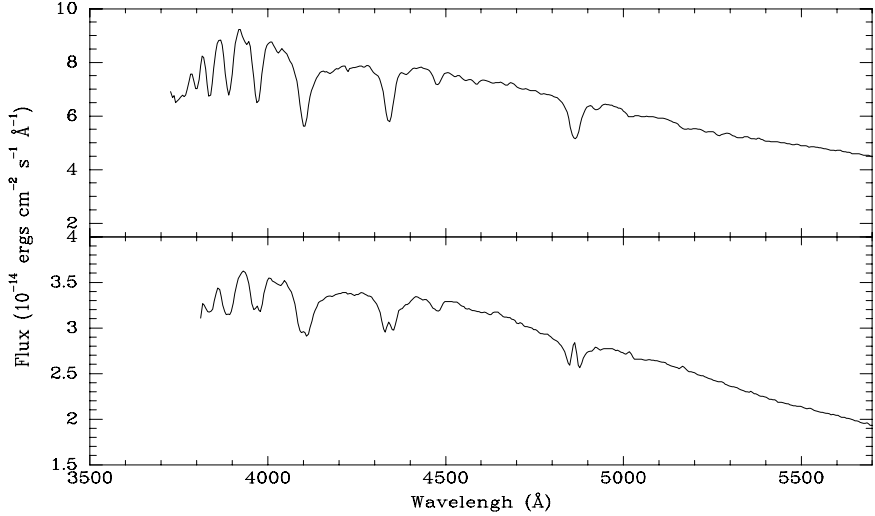}
\caption{Spectra of YZ Cnc during superoutburst. The top spectrum,
obtained on January 21, is much bluer and the flux is also much
higher than the bottom one, obtained on January 22. \label{fig1}}
\end{figure}

%--------------------Table 2-------------------------------
\begin{table}
\caption{Equivalent widths of spectral lines.}
\begin{tabular}{ccccc}
\hline\hline
 Date  & Element           & EW         & Element               & EW\\
(2003) &Rest Wavelength  & (\AA)    & Rest Wavelength         &(\AA)\\
\hline
\ &H$\zeta$ $\lambda$3889&-6.2 &H$\beta$ $\lambda$4861&-8.5\\
\ &H$\epsilon$ $\lambda$3970 &-6.9  &He~I $\lambda$4471 &-1.7\\
Jan 21 &H$\delta$ $\lambda$4101 &-11.3 &He~I $\lambda$4922& -0.9\\
\ &H$\gamma$ $\lambda$4340 &-8.2 &He~I $\lambda$5015+Fe~II $\lambda$5018 &-1.7\\
\hline
\ &H$\zeta$ $\lambda$3889&-4.2 &H$\beta$ $\lambda$4861 (emission)&1.3\\
\ &H$\epsilon$ $\lambda$3970 &-4.7  &H$\beta$ $\lambda$4861& -5.2\\
Jan 22 &H$\delta$ $\lambda$4101 &-8.9 &He~I $\lambda$4471 &-1.3\\
\ &H$\gamma$ $\lambda$4340 &-7.6 &Fe~II $\lambda$5169 (emission) & 0.18\\
\hline
\end{tabular}
\end{table}

\subsection{Radial Velocity}

In Fig.~2A we show the normalized spectrum which is the sum of all
15 individual spectra obtained on January 21. And Fig.~2B shows
the normalized spectrum obtained on January 22. When combined, no
radial velocity shift was applied.

We measured the centers of H$\beta$ absorption troughs of January
21 and emission peaks of January 22 with Gaussian-fit method. We
used H$\beta$ line because it had good signal-to-noise ratios in
both nights. Fig.~3 shows the velocities folded on the orbital
period with the best-fit sinusoidal curve superposed. The orbital
phase was computed according to the ephemeris given by SH,
\[T_0=HJD 2,446,113.794+0.0868(2)E\]
where $T_0$ is the time of the $\gamma$ crossover from negative to
positive velocities and $E$ is a cycle number. The best-fit
sinusoidal shows that H$\beta$ has an amplitude, $K$, of 49$\pm$10
km~ s$^{-1}$ and a systemic velocity, $\gamma$, of 62$\pm$7 km
s$^{-1}$. The value of $K$ is very consistent with the result of
SH, 50$\pm$20 km s$^{-1}$, which was measured during the
quiescence. But the $\gamma$ velocity is somewhat larger than the
value of $\sim$16 km s$^{-1}$ in SH.

It is clearly shown in Fig.~3 that there are two abnormal points
near phase 0.6 and 1 whose velocities are much larger than the
others'. This maybe occur at the accretion flow.

%-------------------------Figure 2-----------------------------
\begin{figure}
\centering
\includegraphics[width=13cm]{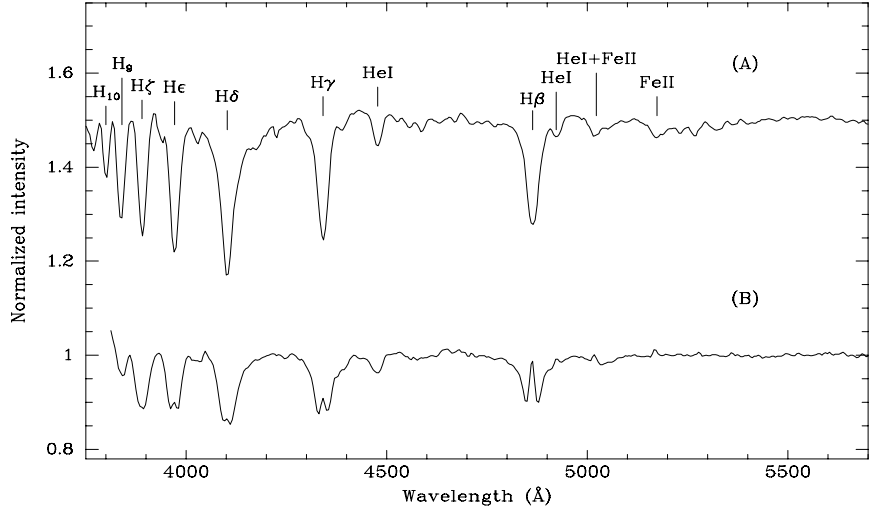}
\caption{Normalized average spectra of YZ Cnc during
superoutburst. (A):observed on January 21; there was no emission
component in \emph{all} spectral lines. (B): observed on January
22; almost all Balmer absorptions were partially filled by
emission on this day. The emission component also came out in He~I
$\lambda$5015+Fe~II $\lambda$5018 absorption. Moreover, the Fe~II
$\lambda$5169 had gone into emission. These differences showed
that the star was going back to quiescence. \label{fig2}}
\end{figure}

%----------------------Figure 3---------------------------------
\begin{figure}
\centering
\includegraphics[width=13cm]{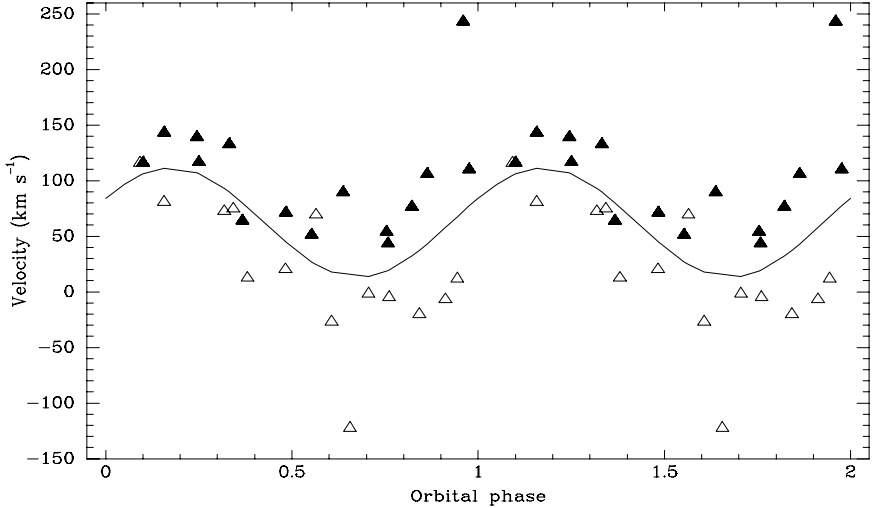}
\caption{Least-squares sinusoidal fitted for the radial velocities
of the centers of H$\beta$ obtained on January 21 and 22,
representing with filled and open triangles respectively. Note
that almost all velocities in the first night are larger than
those in the second night. And it clearly shows that there are two
abnormal points near phase 0.6 and 1 whose velocities are much
larger than the others'. This maybe occur at the accretion flow.
\label{fig3}}
\end{figure}

\subsection{An Eccentric Disk}
\subsubsection{The variation of $\gamma$ and orbital-averaged velocity}
In Fig.~3, we show the radial velocities marked with filled and
open triangles, corresponding to January 21 and 22, respectively.
It can be seen clearly that there is a systemic discrepancy
between these velocities obtained in these two days. We have used
the sky emission line 5577\AA\ to check whether this occurred due
to the systemic error and the difference between these two days is
less than 0.2\AA.  So we believe the existence of discrepancy
between these two $\gamma$ velocities is real. And we derived
these two $\gamma$ velocities from fitting sinusoidal to the data
shown in Fig.~3 (excluding the two abnormal points near the phase
of 0.6 and 1.0, also see Fig.~4) as 91 km s$^{-1}$ and 34 km
s$^{-1}$, for January 21 and 22 respectively.

We also measured the centers of H$\beta$ of the average spectra of
these two days. And we obtained that the relative velocities,
which are averaged throughout the orbital period, are 97 km
s$^{-1}$ and 28 km s$^{-1}$, corresponding to January 21 and 22,
respectively. This phenomenon was also found in IY UMa (Wu et al.
2001), KS UMa (Zhao et al. 2005a). These two features of our
radial velocities are summarized in Table 3.

\subsubsection{The large phase shift of $\sim$0.09 between two days}
As shown in Fig.~4, there obviously existed a phase shift of
$\sim$0.09 between the two sinusoidal velocity curves of January
21 and January 22, as shown by continuous and dashed line
respectively. This feature of the radial velocities is discovered
for the first time for YZ Cnc, even for SU UMa stars. Such big
phase shift in one day could not be due to the uncertainty of the
orbital period because the error of the orbital period of
$2\times10^{-4}$ day given by SH only gives a phase shift of
$\sim0.026$, which is in agreement with the phase offset on
January 22. So there must exist some other reasons responding for
this phenomenon. The feature of phase shift of our radial velocity
curve is also listed in Table 3.

%---------------------------Table 3-----------------------
\begin{table}
\caption{The $\gamma$ velocity, $K$, phase offset and
orbital-averaged velocity}
\begin{tabular}{ccccc}
\hline\hline
 Date (UT) & $\gamma$  &$K$ & Phase offset & $V_{average}$\\
(Year 2002)&  (km s$^{-1}$)& (km s$^{-1}$)&  &(km s$^{-1}$) \\
 \hline
Jan 21 ........& 91$\pm$4 &46$\pm$5 & -0.12$\pm$0.02 & 97$\pm$12\\
Jan 22 ........& 34$\pm$6 &49$\pm$9& -0.03$\pm$0.03 & 28$\pm$10\\
\hline
Shift & 57$\pm$7   & ...& 0.09$\pm$0.04 &69$\pm$16\\
\hline
Jan 21 \& 22  &62$\pm7$&49$\pm$10&-0.07$\pm$04 &...\\
\hline
\end{tabular}
\end{table}

%--------------------------Figure 4--------------------------
\begin{figure}
\centering
\includegraphics[width=13cm]{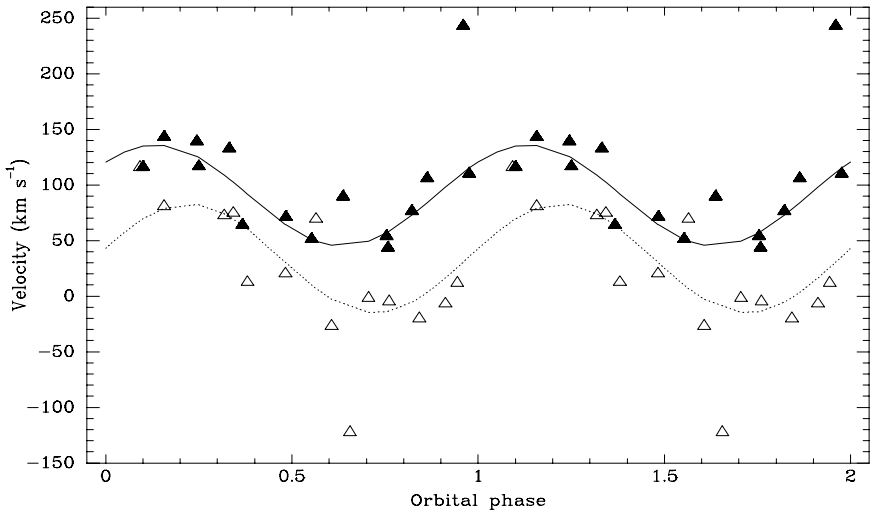}
\caption{Phase shifted in two days. The least-squares sinusoidal
fitted results for the radial velocities on January 21 (filled
triangles) and 22 (open triangles) are present with solid line and
dashed line, respectively. It is obviously that there existed a
shift ($\sim$0.09) between the phases in these two days.
\label{fig4}}
\end{figure}

\subsubsection{An eccentric disk}
The phenomena described above can be well explained with a slow
precession of an eccentric outer disk. The relative velocity
(line-of-sight component) at point $r(\theta)$ on the boundary to
white dwarf is (Wu et al. 2001)
\begin{equation}
V(r,\theta)=C[-e \sin (\theta_0)- \sin(\theta+\theta_0)]
\end{equation}
where $C=\sin i \sqrt{\frac{GM_1}{a(1-e^2)}}=constant$; $i$, $a$
and $e$ are the inclination, half of the major axis and
eccentricity of the accretion disk, respectively. Hence, the mean
velocities of the troughs of the absorption lines or the peaks of
the emission lines are $V=-Ce\sin(\theta_0)$. According to this
equation, the increase of $\theta_0$ will lead to the result that
the central wavelengths of spectral lines, i.e., the
orbital-averaged velocity, will be variable with the precessing
phase of the disk.

Generally, the $\gamma$ velocity is thought to represent the
systemic motion of the binary. This is correct provided that the
accretion disk was axis-symmetry. If the material in the ring
surrounds the primary is not in circular orbits, i.e., the
accretion disk is eccentric and processing, then white dwarf will
be at one of the focus and the mass center of the system (nearly
locating at the geometry center) would change with the precession
period of the disk. Thus, the systemic velocity, i.e., the
$\gamma$ velocity, maybe also change with different precession
phase of the disk.

The large phase shift of $\sim$0.09 between these two days shows
that the variation of the centers of the absorption or emission
peaks can not represent the movement of the white dwarf. And this
phenomenon can also been interpreted with a slow precession of an
eccentric disk. As described above, the position of the mass
center of the system will vary with the precession of the disk.
This would lead to the variation of the time ($T_0$) of the
$\gamma$ crossover from negative to positive velocities, resulting
in the phase shift with the disk at different precession phase.

\subsubsection{A constraint on the eccentricity}
The precessing period ($P_{prec}$) of the disk is the beat period
of the orbital period ($P_{orb}$) and the superhump period
($P_{sh}$). It can be written as
\begin{equation}
\frac{1}{P_{prec}}=\frac{1}{P_{orb}}-\frac{1}{P_{sh}}
\end{equation}
According to Eq. (2), the precessing period of YZ Cnc is 1.52
days, computed with $P_{orb}$=0.0868 day, given by SH and
$P_{sh}$=0.09204, given by Patterson (1979). Thus, $\theta_0$ will
increase 4.12 rad within 1 day. So we have the mean velocities of
the absorption troughs to be $-Ce\sin(\theta_0)$ on January 21,
the emission peaks to be $-Ce\sin(\theta_0+4.12)$ on January 22.
Comparing their difference with the measured data (the shift of
``$V_{average}$'' in Table 3), we have
\begin{equation}
C e \cos(\theta_0+2.06)=60
\end{equation}
If we know the value of $C$, i.e., know $M_1$, $a$, $q$ and
$P_{orb}$, we can give a constraint on the eccentricity of the
disk.
\subsection{Mass and Inclination}
The mass and inclination of stars are very difficult to be
determined if the system is not eclipsing. SH used the properties
of the emission lines (specially the linewidth) to provide a
relation between the inclination and the white dwarf mass with an
empirical assumption that the velocity from the line center at
30\% of the emission-line intensity best represents $V_d\sin i$.
They gave that the mass of the secondary is $\sim$0.17 $M_\odot$
and 0.75-0.9 $M_\odot$ of the primary.

Substituting $P_{orb}$ with 0.0868(2) day in equation (1) in Zhao
et al. (2005a), we have
\begin{equation}
M_1=M_2/q, \ M_2=[0.829(1+1/q)^{1/3}Q(q)]^{15/7}
\end{equation}
We can obtain the mass ratio by using an empirical relation found
by Patterson (2001), $\epsilon=0.216(\pm0.018)q$, where
$\epsilon=(P_{sh}-P_{orb})/P_{orb}$. It gives $q=0.28\pm0.03$,
which is somewhat larger than that of SH.

According to equation (4), it is only requires $q>0.09$ to meet
the condition that the white dwarf mass should be less than 1.44
$M_{\odot}$. The mass ratio of 0.28$\pm0.03$ derived above is
consistent with this requirement. Therefore we can obtain that
$M_{2}=0.13\pm0.01\ M_{\odot}$ and $M_{1}=0.46\pm0.06M_{\odot}$.
Using the $K_1=49\pm10$ km s${^-1}$ (see \S3.2), the mass function
$f(M)=(M_{2}\sin i)^3/(M_1+M_2)^2]=K_1^3P_{orb}/(2\pi
G)=0.00106(10)\ M_{\odot}$ gives $i=34^\circ\pm9^\circ$.

The values of $M_1$ and $M_2$ are smaller than those of SH. This
is believed due to the systemic difference between different
methods and some unproved empirical assumptions. Hence, the masses
and inclination given here are rather uncertain.

\section{Discussion}

The variation of $\gamma$ velocity discovered in YZ~Cnc is for the
first time but not alone. It has been discovered in several other
SU~UMa stars, i.e., Z~Cha (Vogt 1982, Honey et al., 1988), KS~UMa
(Zhao et al. 2005a) and ER~UMa (Zhao et al. 2005b). As described
by these different groups, the RV curves for different CVs are
\emph{all} gotten by measuring spectra obtained when the stars
went through \emph{eruptions}.

Vogt (1982) and Honey et al. (1988) have found that the $\gamma$
velocities of Z~Cha varied during its superoutburst. Vogt (1981)
proposed a model in which he considered the behavior of a
precessing, elliptical ring surrounding a circular accretion disk.
This gives the variation of the $\gamma$ velocity on a
night-to-night basis as a result of variations in the projected
motion of the ring material against that of the inner (circular)
disk. Honey et al. (1988) interpreted their observational result
with new non-axisymmetric disk simulations as arising in an
eccentric, precessing disk which is tidally distorted by the
secondary.

Our results that the $\gamma$ velocity vary with time and that the
phase shifts between different days are based on the measurement
of centers of the absorption troughs of H$\beta$ and the centers
of the peaks of the emission cores. We can not help but do this.
We can't use the double-Gaussian convolution method (Shafter et
al. 1988) to measure RV because the wings of H$\beta$ are blended
with He I $\lambda$4922 and the other Balmer lines not only are
contaminated but also have bad signal-to-noise ratio, especially
on January 22. Despite of this, our observational result confirmed
that the $\gamma$ velocity do actually vary when the star was
ongoing a superoutburst.

We find for the first time that the phase of the system would
change. If we believe that the orbital period and the error given
by SH are reliable, the phase shift between these 2 days does
actually exist. It is not surprised that this phenomenon can be
found in YZ Cnc between 2 days because the precession period of
YZ~Cnc is only 1.52 days. The phase shift is enough to be observed
between 2 days. So our observation provides more evidences to
convince us that the accretion disk is precessing and eccentric
when the binary system going through superoutburst. If the system
has a larger inclination, we could get more information and do
more detailed analysis like IY UMa (Wu et al. 2001) and KS UMa
(Zhao et al. 2005a).

We can also estimate the eccentricity crudely by substituting the
mass of the white dwarf $M_1$, the mass ratio $q$ and the
inclination $i$, with the values given above (see $\S$3.4), we
obtained
\[\cos(\theta_0+2.06)=0.231/e\]
If we substituted these parameter with the values given by SH, we
would get
\[\cos(\theta_0+2.06)=0.184/e\]
So $e$ must be larger than 0.184 or 0.231, according to whose
parameters are more reliable.

\section{Conclusions}
So far, we have shown the properties of our spectra and the radial
velocities of YZ Cnc obtained during its 2002 January
superoutburst. These properties include following three aspects,

(1) The $\gamma$ velocity (62$\pm$7 km s$^{-1}$) obtained in these
two days deviated strongly from the systemic velocity (16$\pm$10
km s$^{-1}$) measured by SH when the binary system was in
quiescence. And there is a discrepancy of $\sim$60 km s$^{-1}$ of
the $\gamma$ velocities between these two days.

(2) The mean velocities averaged throughout the orbital period of
these two days have large offset of the order of $\pm$70 km
s$^{-1}$.

(3) There is large phase offset of $\sim$0.09 between these two
days.

As detailedly described in $\S$3.3, we can make a conclusion that
these features are all ascribing to the precession of an eccentric
accretion disk. Therefore, we can make use of these properties to
confirm whether the accretion disk is eccentric or not.

\begin{acknowledgements}
We would like to thank the Optical Astronomy Laboratory, Chinese
Academy of Sciences and Prof. Jianyan Wei of the National
Astronomical Observatory for scheduling the observations. We are
grateful for the financial support from the National Natural
Science Foundation of China (through grants 10173005 and
10010120074).
\end{acknowledgements}

\end{document}